\documentclass[11pt]{openjournal}
\usepackage{tabularx}
\usepackage{graphicx}
\usepackage{cancel}
\usepackage{amssymb}

\usepackage{amsmath}

\usepackage{color}
\usepackage[pdftex, colorlinks=true, linkcolor=black, urlcolor=black,  citecolor=blue]{hyperref}
\usepackage{subfigure} 
\usepackage{float}

\def\MPl{M_{\rm{Pl}}}

\def\half{\frac{1}{2}}
\def\adot{\dot{a}}
\def\phidot{\dot{\phi}}
\def\wphi{w_{\phi}}
\def\weff{w_{\rm{eff}}}
\def\wM{w_{\rm{M}}}
\def\wm{w_{\rm{m}}}
\def\wr{w_{\rm{r}}}
\def\Omp{\Omega_{\phi}}
\def\OmK{\Omega_{K}}
\def\Omm{\Omega_{\rm{m}}}
\def\OmM{\Omega_{\rm{M}}}
\def\Omr{\Omega_{\rm{r}}}
\def\rhom{\rho_{\rm{m}}}
\def\rhor{\rho_{\rm{r}}}
\def\rhoM{\rho_{\rm{M}}}
\def\PM{P_{\rm{M}}}
\def\HoverH{\frac{\dot{H}}{H^2}}

\begin{document}
\title{Dynamical Analysis of Scalar Field Cosmologies with Spatial Curvature}

\author{Mateja Gosenca}
\author{Peter Coles}
\email{M.Gosenca@sussex.ac.uk}
\email{P.Coles@sussex.ac.uk}
\address{Astronomy Centre, School of Mathematical and Physical Sciences, University of
Sussex, Brighton BN1 9QH, United Kingdom}

\begin{abstract}
\noindent
We explore the dynamical behaviour of cosmological models involving
a scalar field (with an exponential potential and a canonical
kinetic term) and a matter fluid with spatial curvature included in
the equations of motion. Using appropriately defined parameters to
describe the evolution of the scalar field energy in this situation,
we elucidate the character of two fixed points that are not present
in the case without curvature. We also analyse the evolution
of the effective equation-of-state parameter for different initial
values of the curvature.
\end{abstract}

\maketitle

\section{Introduction}
\noindent
The discovery, based on the observed behaviour of Type Ia Supernovae
\cite{astro-ph/9805201, astro-ph/9812133}, that the expansion of the Universe
appears to be accelerating, generated enormous theoretical
interest in finding a suitable framework to account for this
phenomenon. Other independent observational data such as the Cosmic
Microwave Background radiation (CMB) \cite{1212.5226, 1303.5076} and Baryonic Acoustic Oscillations (BAO)
\cite {0907.1660, 0812.0649} have subsequently confirmed that reconciling a standard
Friedmann-Lema\^{i}tre-Robertson-Walker model based on Einstein's
General Theory of Relativity requires roughly $70\%$ of the
average energy density today to be in the form of an exotic fluid
whose equation-of-state parameter $w=P/\rho$ is close to $-1$.

Although over fifteen years have passed since the original
discovery, we still lack a compelling theoretical model to explain
this so-called dark energy. The three major theories that attempt to
do this all rely in one way or another on modification of the
Einstein Equations, which we use in the form
\begin{equation} R_{\mu \nu} - R g_{\mu \nu}/2
=8\pi G T_{\mu \nu}, \end{equation} in which properties of
space-time appear on the left-hand side and the matter-energy
contents of the Universe on the right.

The obvious phenomenological possibility is to add a constant term
$\Lambda g_{\mu \nu}$ to Einstein's Equations. This seems appealing
because of its simplicity and that when included in the
energy-momentum tensor on the right-hand-side of the Einstein
Equations, it leads to a clear prediction of the effective equation
of state, namely that $w=-1$ exactly. There are, however, two major
concerns with this picture. The first is that the magnitude of the
vacuum energy associated with a $\Lambda$-term is out of line with
the value expected from summation of zero-point energies of quantum fields up to some cut-off scale. This discrepancy is 120 orders of magnitude if one chooses a cut-off at the Planck energy, but even if some mechanism imposes a cut-off at for example a QCD energy scale, the problem is alleviated but not entirely solved \cite{RevModPhys.61.1}.

The second problem is that there seems no natural
explanation of why the energy density associated with the vacuum
energy should be within a factor of a few of the present matter
density or, in other words, why the expansion of the Universe should
have began to accelerate so very recently in cosmic history. These
are called the fine-tuning problem and the coincidence problem,
respectively \cite{1212.4726}. Although such statements clearly depend on some choice of measure, related in this case to the probability distribution of the value of $\Lambda$, they must also take account of anthropic selection effects \cite{Sivanandam:2012ty}.

A second class of models that might explain dark energy are those
based on some form of modification of Einstein's theory of gravity.
The usual approach in such models is that instead of starting from
the standard Einstein-Hilbert action (which leads to the Einstein
Equations), one considers additional terms in the action
\cite{1101.0191}. The most straightforward modification of
General Relativity is to replace the Ricci scalar, $R$, in the
Einstein-Hilbert action, by ther function of this scalar, usually
called $f(R)$. An example of this model is $f(R)= R+\alpha R^2$,
which was in fact one of the first models of inflation, proposed by
Starobinsky. The idea of using $f(R)$ for late-time acceleration was
first suggested in \cite{gr-qc/0201033} and examples of viable
models of this type are proposed in \cite{0706.2041, 0708.1190,
0709.1391}. In a more general case, one can include an arbitrary
function of $R$, $R_{\mu \nu} R^{\mu \nu}$ and $R_{\mu \nu \rho
\sigma} R^{\mu \nu \rho \sigma}$ in the action. The Gauss-Bonnet
combination $ R^2 - 4 R_{\mu \nu} R^{\mu \nu} + R_{\mu \nu \rho
\sigma} R^{\mu \nu \rho \sigma}$ \cite{hep-th/0411102} is particularly
widely explored in the literature. Another, more complicated, class
of modified gravity models include scalar-tensor theories, where
Ricci scalar $R$ and a scalar field $\phi$ are coupled (an example
of this is Brans-Dicke theory \cite{Brans:1961sx}), and DGP (Dvali,
Gabadadze and Porrati) braneworld model \cite{hep-th/0005016}. In the
latter, particles are confined to a 3-dimensional brane, embedded in
a 5(or more)-dimensional space-time with an infinite extra
dimension. Standard 4D gravity is recovered at small distances, but
at larger scales gravity is weakened, because its energy is
essentially getting lost to the additional dimension.

A third class of possible models include modifications to the
standard form of matter on the right-hand-side of the Einstein
Equations, designed to generate a negative
effective pressure. Among such models are quintessence, $k$-essence,
coupled dark energy and the generalised Chaplygin gas (for reviews
see \cite{1004.1493, 1304.1961, 1212.4726, hep-th/0603057}).
Quintessence, which will be studied in much greater detail in the
rest of this paper, represents the idea that accelerated expansion
is driven by a canonical scalar field $\phi$ \cite{Wetterich1988,
astro-ph/9708069, astro-ph/9807002}. The most important consequence of this is
that we now have a dynamical equation of state, rather than a
constant \cite{gr-qc/9711068, hep-th/0603057}. Please note that this is also the case in some of the modified-gravity theories discussed in the paragraph above. Quintessence models are
usually divided into two types on the basis of the form of potential
that drives the scalar field dynamics. The first class contains
models in which $w$ gradually decreases to $w=-1$; these models are
called ``freezing models'' \cite{astro-ph/0505494}. The forms of
potential needed for this kind of behaviour were studied in
\cite{astro-ph/9812313, PhysRevD.37.3406}. In this scenario, the field
energy density does not necessarily need to be negligible at the
radiation epoch (this is the case with the cosmological constant).
The other option is the class of ``thawing models''
\cite{astro-ph/0505494} in which the field equation-of-state parameter
is close to $w=-1$ initially, but at late times deviates from this
value.

Theories that involve non-canonical kinetic terms in the Lagrangian
are called $k$-essence \cite{astro-ph/9912463, astro-ph/0006373,
1401.6339}. The idea is that inclusion of non-canonical terms
results in cosmic acceleration even without the field potential, as
was shown in \cite{hep-th/9904075}. A number of different
scenarios were proposed, to name a few: Low energy effective string
theory \cite{hep-th/9211021}, Ghost condensate \cite{hep-th/0312099},
Tachyon field \cite{hep-th/0003122}, Dirac-Born-Infeld (DBI) theory
\cite{hep-th/0310221, hep-th/0404084}.

Furthermore, there are suggestions that dark energy and dark matter are coupled \cite{hep-th/9408025, astro-ph/0702015, astro-ph/0309411}, that a single fluid (e.g. generalised Chaplygin gas) is responsible for dark energy and dark matter \cite{gr-qc/0103004, astro-ph/0210468}, and finally unified quintessence and inflation theories, namely quintessential inflation
\cite{astro-ph/9810509, astro-ph/0605205}. In \cite{arXiv:1501.06540} the authors consider models of quintessence interacting with dark or baryonic matter.

There are also attempts to explain the apparent cosmic acceleration
by means of inhomogeneities in the matter distribution in other
words by breaking the assumption that cosmological space-time is
described by a FLRW metric; for a review see \cite{0707.2153}. One
specific example is the suggestion that we live in the middle of a
huge underdensity (a ``void'') and we interpret the expansion of its
surroundings as an overall cosmic acceleration
\cite{astro-ph/0011484, astro-ph/0512006}, although this seems to be in conflict with observations, see for example \cite{Bull:2011wi}. Another approach relies
upon the back-reaction of cosmological perturbations
\cite{astro-ph/0311257, hep-ph/0409038}, which may be able to
explain acceleration without dark energy; some work along these
lines related to this paper can be found in \cite{gr-qc/0606020,
gr-qc/1103.1146}.

Given this plethora of possible
models it is important to undertake a rigorous systematic study of their dynamical properties, in order to understand and classify the wide range of behaviours they may exhibit. In particular, focusing on the fixed points of their evolution will allow us to tackle the difficult question of what can be considered to be ``natural'' behaviour in a given scenario. Since the fixed points attract trajectories from a wider parameter space around them, the configuration of the system can be determined by its dynamical evolution in the fixed point rather than in the initial conditions. We concentrate on a specific class of possibilities by studying the dynamical evolution of quintessence-type
models, extending previous work (described in detail below) by
including spatial curvature, which is usually neglected in such
analyses. Although the observational evidence at the moment points to a universe
which is (nearly) spatially flat this conclusion is based on a model with restricted set of parameters. In the framework of a more general model the constraints on $\OmK$ can be weaker \cite{Okouma:2012dy}.
It therefore remains important to establish whether the inclusion of curvature leads to any qualitative changes in the dynamics. This issue has been addressed before \cite{gr-qc/9901014}, but in this paper we use a different parametrisation of the equations of motion, which makes the connection between the curvature and dynamics more explicit. The aim of our analysis is to ask the question whether
it is possible, via this relatively simple generalisation of the
quintessence scenario, to generate an attractor solution that
corresponds to a value of $w\neq -1$, which would distinguish it
from many of the other models listed above. 

The paper is organized as follows.
In Sec. \ref{sec:analysis} we explore the dynamics of a single scalar field with an exponential potential and a canonical kinetic term under the flat FLRW metric.
This is a well-studied system \cite{hall87,burdbar88} but one which
repays further analysis. We find the fixed points of this system and
investigate values of important dynamical parameters at these fixed
points; this section follows the analysis that was done in
\cite{gr-qc/9711068}. The further details of this analysis are given in the Appendix.
In Sec. \ref{sec:curvature} we generalize this
approach by introducing another variable that corresponds to spatial
curvature and analyze how the dynamics are affected as a
consequence. We find two new fixed points and explore evolution of
equation-of-state parameter for different initial values of
curvature. For a parallel discussion see \cite{astro-ph/1307.7399}.

\section{Dynamical analysis}
\label{sec:analysis}
\subsection{Background}
\noindent
The action of a canonical scalar field $\phi$ is given by
\begin{equation}
S = \int d^4 x \sqrt{-g} \left[ \frac{1}{2}\MPl^2 R - \frac{1}{2} g^{\mu \nu} \partial_{\mu}\phi \partial_{\nu}\phi -V(\phi) + \mathcal{L}_{\rm{M}} \right] ,
\end{equation}
where $\MPl$ represents the reduced Plank mass: $\MPl = 1/\sqrt{8\pi G} = 1/\sqrt{\kappa} $. The matter Lagrangian $\mathcal{L}_{\rm{M}}$ has two contributions: non-relativistic matter with equation of state $\wm=P_m/\rhom=0$ and radiation with equation of state $\wr=1/3$. Varying this action with respect to metric and applying the action principle gives the Einstein Equations with the energy momentum tensor:
\begin{equation}
T_{\mu \nu} =  \partial_{\mu} \phi \partial_{\nu} \phi - g_{\mu\nu} \left( \frac{1}{2} \partial_{\rho} \phi \partial^{\rho} \phi -V(\phi) \right).
\label{eq:tensor:scalar}
\end{equation}
The assumption of spatial homogeneity and isotropy allows us to adopt the Friedmann-Lema\^{i}tre-Robertson-Walker (FLRW) metric:
\begin{equation}
ds^2=-dt^2 + a(t)^2 \left( \frac{dr^2}{1-Kr^2} + r^2 d\theta^2 +r^2\sin{\theta}^2 d\phi^2 \right),
\end{equation}
where $a(t)$ is time-dependent scale factor and K is a constant describing the spatial curvature. The universe is open if $K<0$, flat if $K=0$ and closed if $K>0$. With the use of this metric, the Einstein Equations become the modified Friedmann Equations: 
\begin{subequations}\label{eq:Fridman}
\begin{equation}\label{eq:Fridman:scalar:one}
3\MPl^2 H^2 = \half \dot{\phi}^2 +V(\phi)+\rhom + \rhor  - 3 \MPl^2 \frac{K}{a^2}
\end{equation}
\begin{equation}\label{eq:Fridman:scalar:two} 
2\MPl^2 \dot{H} = - \dot{\phi}^2 - (1+\wm)\rhom - (1+\wr)\rhor  +2 \MPl^2 \frac{K}{a^2}
\end{equation}
\end{subequations}
where the Hubble parameter is defined in the usual manner as
$H=\adot/a$. The requirement of energy-momentum conservation demands
$T^{\mu \nu}_{\;\;\;\; ; \nu}=0$. For a homogeneous and isotropic
universe the energy-momentum tensor becomes symmetric, i.e. $T_{\mu
\nu}=\mathrm{diag}(-\rho, P, P, P)$ and we thus obtain the
continuity equation $\dot{\rho}+3H(\rho +P)=0$. This equation is
obeyed separately by the matter, radiation and scalar field as long as there is no coupling between these components.  Applying the
same condition to the energy-momentum tensor of the scalar field
(\ref{eq:tensor:scalar}) we get
\begin{equation}
\ddot{\phi} + 3H\dot{\phi} + V_{,\phi} =0\; ,
\label{eq:continuity:scalar}
\end{equation}
where $V_{,\phi}$ is the  derivative of the potential with respect
to the field $\phi$. Comparing this to the general continuity
equation we see that the effective energy density and pressure of
the scalar field are $\rho_\phi=\dot{\phi}^2/2 +V(\phi)$ and
$P_\phi=\dot{\phi}^2/2 -V(\phi)$, respectively. The equation of
state for the scalar field is then given by
\begin{equation}
w_{\phi}=\frac{P_{\phi}}{\rho_{\phi}}=\frac{\dot{\phi}^2/2 -V(\phi)}{\dot{\phi}^2/2 +V(\phi)}.
\end{equation}
Equation (\ref{eq:continuity:scalar}) is a dynamical equation for
the evolution of the scalar field; in order to solve for the
dynamics of the system, one must solve this equation simultaneously with the Friedmann
Equations (\ref{eq:Fridman}) and the continuity equation.

\subsection{Two dynamical variables}
\noindent
It is convenient to define a new set of dimensionless variables:
\begin{equation}\label{eq:variables2D}
x=\frac{\phidot}{\sqrt{6} \MPl H} \;, \hspace{30mm} y=\frac{\sqrt{V(\phi)}}{\sqrt{3} \MPl H}\;,
\end{equation}
because the energy density of the scalar field can be expressed as:
\begin{equation}
\Omega_{\phi}=\frac{\rho_{\phi}}{3 \MPl^2 H^2} = x^2+y^2
\end{equation}
and the equation of state for the scalar field reads:
\begin{equation}
w_{\phi}=\frac{x^2-y^2}{x^2 +y^2}.
\end{equation}
Additionally defining combined matter and radiation energy density parameter as:
\begin{equation}
\OmM=\Omm + \Omr=\frac{\rhom}{3 \MPl^2 H^2}+\frac{\rhor}{3 \MPl^2 H^2}
\end{equation}
and assuming that spatial curvature is zero, the first Friedmann
Equation (\ref{eq:Fridman:scalar:one}) simplifies as:
\begin{equation}\label{eq:constraint}
1=x^2+y^2+\OmM.
\end{equation}
To study the evolution of the scalar field we can take derivatives
of $x$ and $y$ with respect to the number of $e$-foldings
$N=\ln{a}$, anticipating that such a system can display accelerated
exponential expansion. 
For an exponential potential $V=V_0 e^{- \lambda \phi / \MPl}$ we get:
\begin{subequations}\label{eq:derivatives}
 \begin{equation}
\frac{dx}{dN} =\sqrt{\frac{3}{2}} \lambda  y^2-\frac{3}{2} x \left( \wM \left(x^2+y^2-1\right)-x^2+y^2+1 \right)
\end{equation}
\begin{equation}
\frac{dy}{dN} =-\sqrt{\frac{3}{2}} \lambda  xy -\frac{3}{2} y \left( \wM \left(x^2+y^2-1\right)- x^2+ y^2-1 \right).
\end{equation}
\end{subequations}
In order to analyse the dynamical system (\ref{eq:derivatives}) we
first look at the fixed/critical points for different values
of the parameters $\lambda$ (not to be confused with
the cosmological constant $\Lambda$) and $\wM$ (defined as $\wM=(\rhom \wm + \rhor \wr)/(\rhom +\rhor)$). From the constraint (\ref{eq:constraint}), it follows that $x^2 +y^2 \leq 1$ because
$\OmM$ is always positive for the form of potential we use. Every solution with non-zero $y$ can be
positive or negative in $y$, since $y^2 \sim V$. The part of
parameter space in which $y$ is negative corresponds to a
contracting universe. Because the phase plane is symmetric with
respect to the $x$-axis, we will only consider its upper part.
This implies that trajectories in phase plane are limited to the upper half of the unit disc. Fixed points are defined as:
\begin{equation}
\frac{dx}{dN}=0 \;, \hspace{15mm} \frac{dy}{dN}=0\;.
\end{equation}
There are five such critical points; their coordinates and values of
the associated physical parameters are listed in Table
(\ref{tab:2D}).

\begin{table}[ht]
\centering
\caption{Coordinates and properties of fixed points.}
\label{tab:2D}
\begin{tabular}{ |c || c  c || c  c c  |}
\hline
 & $x_*$ & $y_*$  & $\weff$ & $\Omp$& $\OmM$\\
\hline
\hline
A & 0 & 0  & $\wM$&0&1 \\
\hline
$\mathrm{B}_1$  & 1 & 0  & 1 &1&0\\
\hline
$\mathrm{B}_2$  & -1 & 0  & 1 &1&0\\
\hline
D  & $\lambda / \sqrt{6}$ & $\sqrt{1-\lambda ^2 / 6}$ & $\lambda ^2 /3-1$&1&0 \\
\hline
E & $\sqrt{3/2} ( \wM +1) / \lambda $ & $\sqrt{3/2( 1-\wM^2)}/\lambda $ & $\wM$& $3(\wM+1)/\lambda ^2$&$1 - 3(\wM+1)/\lambda ^2$ \\
\hline
\end{tabular}
\end{table}
\noindent
The value of the parameter $\lambda$ can consequently be divided into three qualitatively different cases:\\
\begin{itemize}
\item ($\lambda<\sqrt{3(\wM+1)}$) All trajectories are drawn to the point \textbf{D} which is a \emph{stable} attractor.
\item ($\sqrt{3(\wM+1)}<\lambda<\sqrt{6}$)  Point \textbf{E} becomes a \emph{spiral} attractor. \textbf{D} is still present.
\item ($\sqrt{6}<\lambda$ ) \textbf{D} is no longer defined and $\rm{\textbf{B}}_{\rm{\textbf{1}}}$ becomes a \emph{saddle point}.
\end{itemize}

\begin{figure}[h!]
\centering
\subfigure{
\includegraphics[scale=0.5]{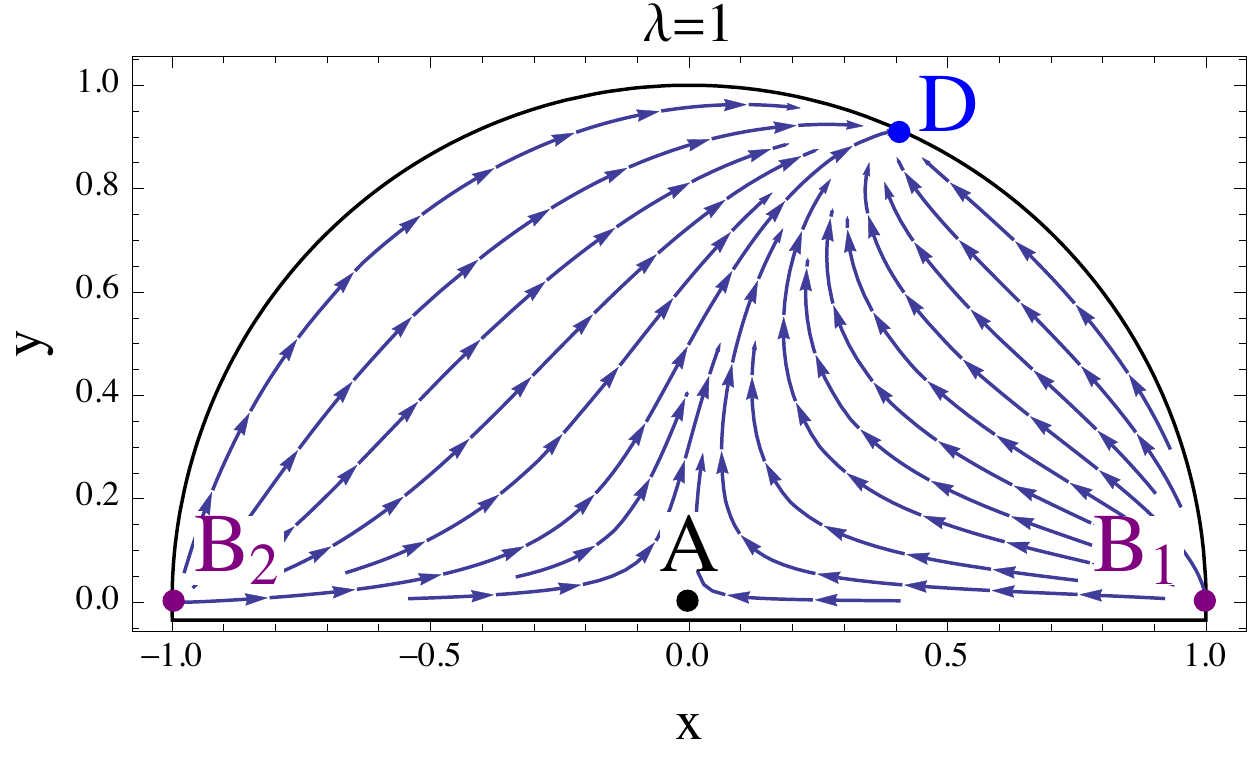}
}
\subfigure{
\includegraphics[scale=0.5]{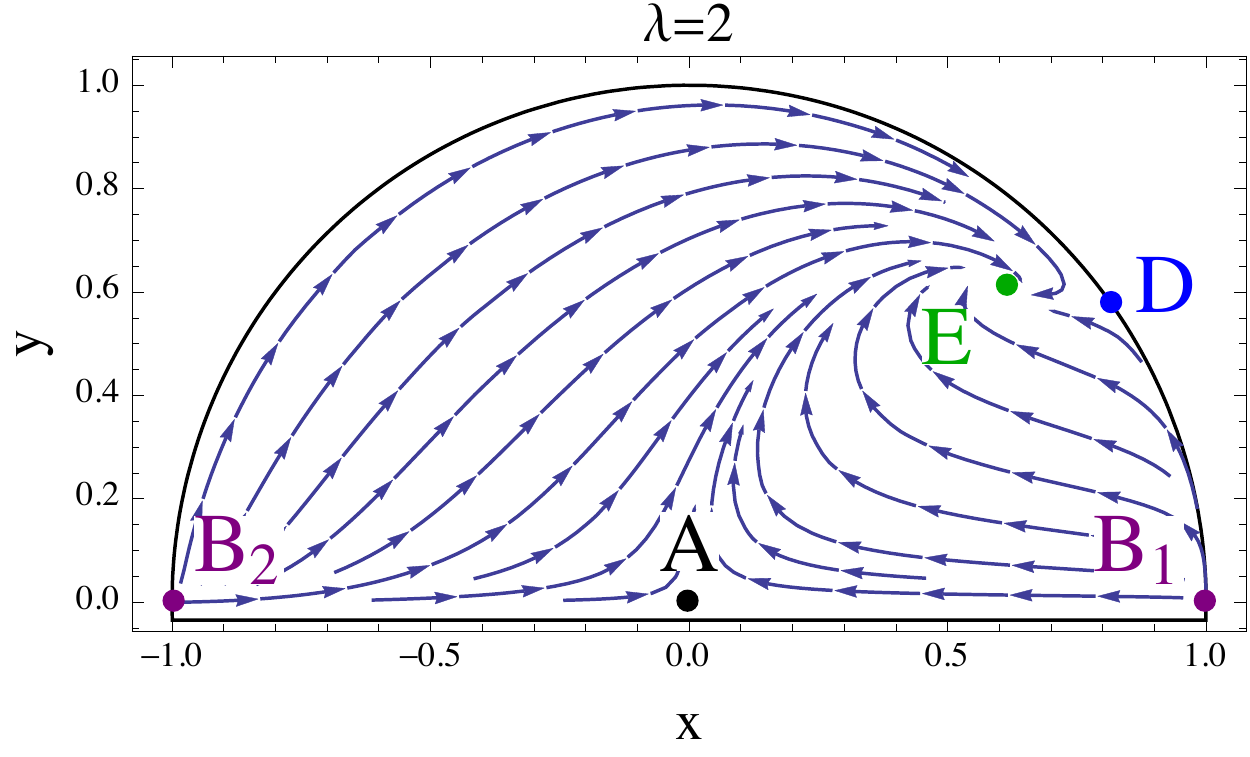}
}
\subfigure{
\includegraphics[scale=0.5]{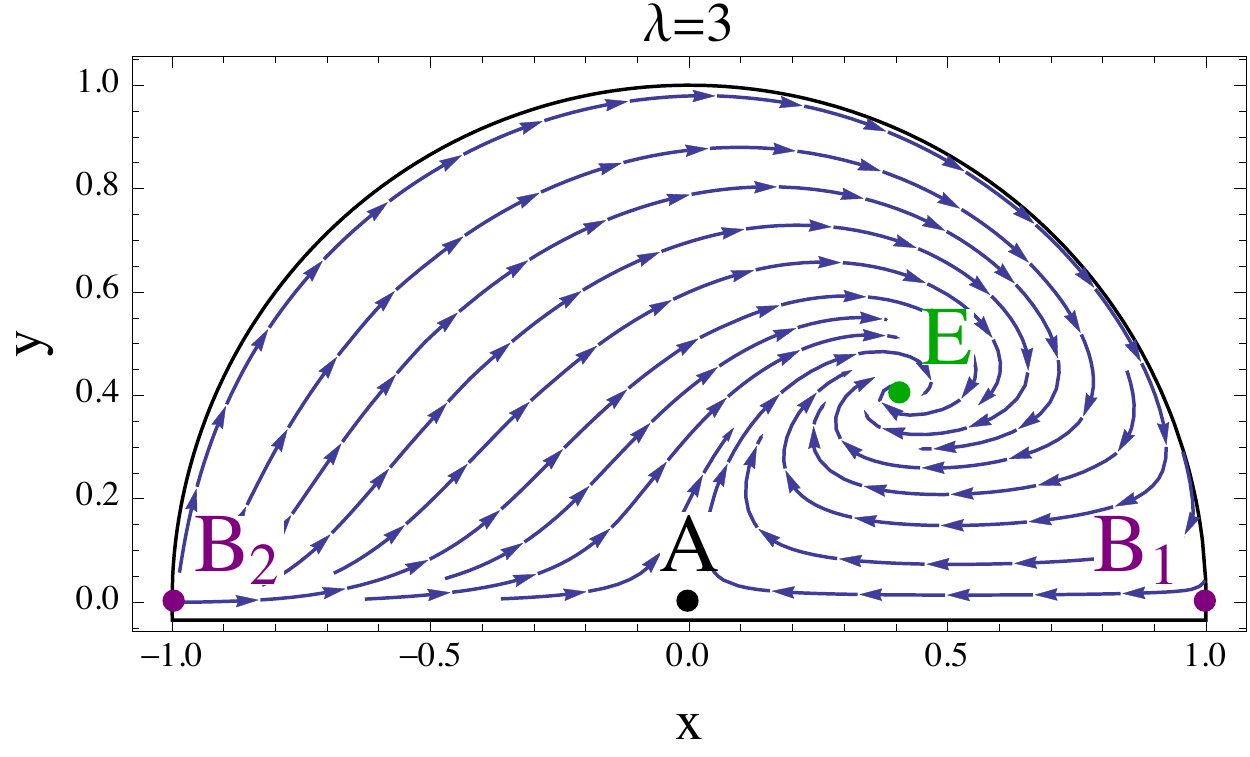}
}
\caption{Phase planes for three qualitatively different cases of $\lambda$ and $\wM=0$ as discussed in the text.  \label{pic_2D_phase_plane}}
\end{figure}

The solutions of Equations (\ref{eq:derivatives}) for these three cases are illustrated in figure (\ref{pic_2D_phase_plane}).
We have defined $\wM$ to be a combination of pressureless dust (with $\wM=0$) and radiation ($\wM=1/3$)
so its values can only lie between these two values but in the most
general case (including more exotic equations of state) its value could lie outside this range.
The specific cases shown in the figures (\ref{pic_2D_phase_plane}) all have $\wM=0$.
The value of this parameter only slightly affects the position of the fixed point \textbf{E} (and none others).
For higher values of  $\wM$ points \textbf{E} and \textbf{D} merge at higher $\lambda$. For a limiting  case $\wM=1$
(that extends beyond our analysis) point \textbf{E} merges with \textbf{D} exactly at the
point $\rm{\textbf{B}}_{\rm{\textbf{1}}}$ so it is  never inside the half disc.  \\

\section{The Role of Curvature}
\subsection{Three dynamical variables}
\label{sec:curvature} In this subsection we introduce spatial
curvature into dynamical system, while still keeping matter and
radiation as one variable and considering a scalar field described
as an exponential potential. We define new variables to construct a
generalization of the two-dimensional phase plane discussed in the
previous subsection into a three-dimensional phase space:
\begin{equation}
x=\frac{\phidot}{\sqrt{6} \MPl  H} \;, \hspace{15mm} y=\frac{\sqrt{V(\phi)}}{\sqrt{3} \MPl  H}\;, \hspace{15mm} z=\frac{K}{a^2 H^2} \;.
\end{equation}
Definitions of $\Omp$, $\wphi$, $\OmM$ and $\wM$ are identical to
those in the previous subsection. The first Friedmann Equation is
now expressed as:
\begin{equation}
1=x^2+y^2-z+\OmM.
\end{equation}
Note that $z$ is not squared in this definition. If $z$ were defined
in such a manner that it was squared in the above equation then all
solutions with negative $K$ would correspond to solutions with
imaginary $z$. The minus sign in front $z$ has no physical
significance; our definition corresponds to $\OmK=-z$. With the
introduction of the curvature, ratio of the Friedmann Equations
becomes:
\begin{equation}
\HoverH= \frac{3}{2} \left(\wM \left(x^2+y^2-z-1\right)- x^2+ y^2-1\right) - \frac{1}{2}z
\end{equation}
and the effective equation-of-state parameter can be expressed as:
\begin{equation}
\weff=\frac{P_{\phi}+\PM}{\rho_{\phi}+\rhoM}=\frac{x^2-y^2+\wM(1-x^2-y^2+z)}{1+z}
\end{equation}
The evolution of the dynamical system is now described by a system
of three coupled differential equations:

\begin{subequations}
\begin{equation}\label{eq:derivatives:3D}
\frac{dx}{dN} =\sqrt{\frac{3}{2}} \lambda  y^2-\frac{1}{2} x \left(3 \wM \left(x^2+y^2-z-1\right)-3 x^2+3 y^2-z+3\right)
\end{equation}
\begin{equation}
\frac{dy}{dN} =\frac{1}{2} y \left(-3 \wM \left(x^2+y^2-z-1\right)+3 x^2-\sqrt{6} \lambda  x-3 y^2+z+3\right)
\end{equation}
\begin{equation}
\frac{dz}{dN} =z \left(-3 \wM \left(x^2+y^2-z-1\right)+3 x^2-3 y^2+z+1\right) .
\end{equation}
\end{subequations}

\subsection{Phase space analysis, fixed points and their stability}
\begin{figure}[h!]
\centering
\subfigure{
\includegraphics[scale=0.53]{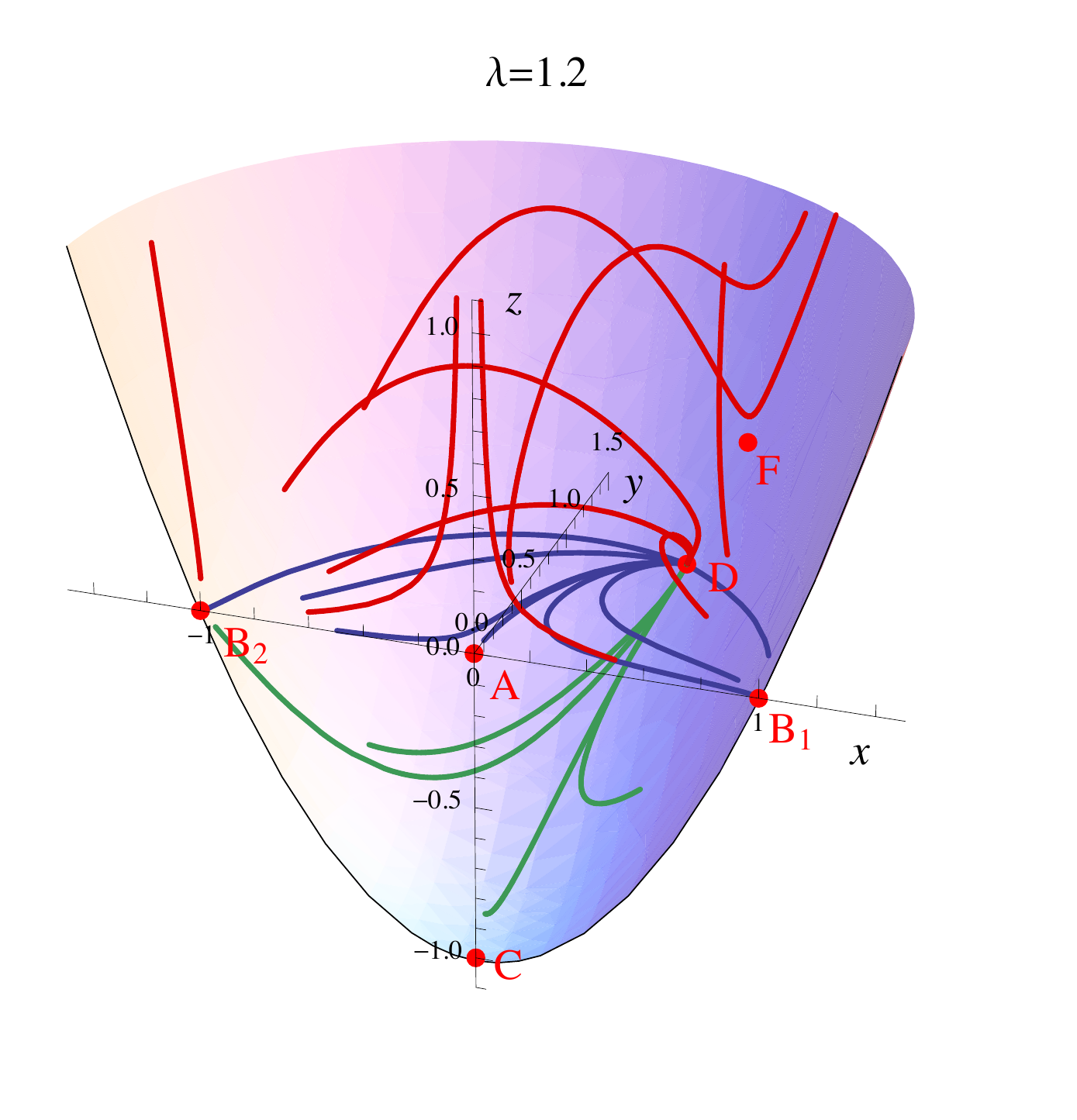}
}
\subfigure{
\includegraphics[scale=0.53]{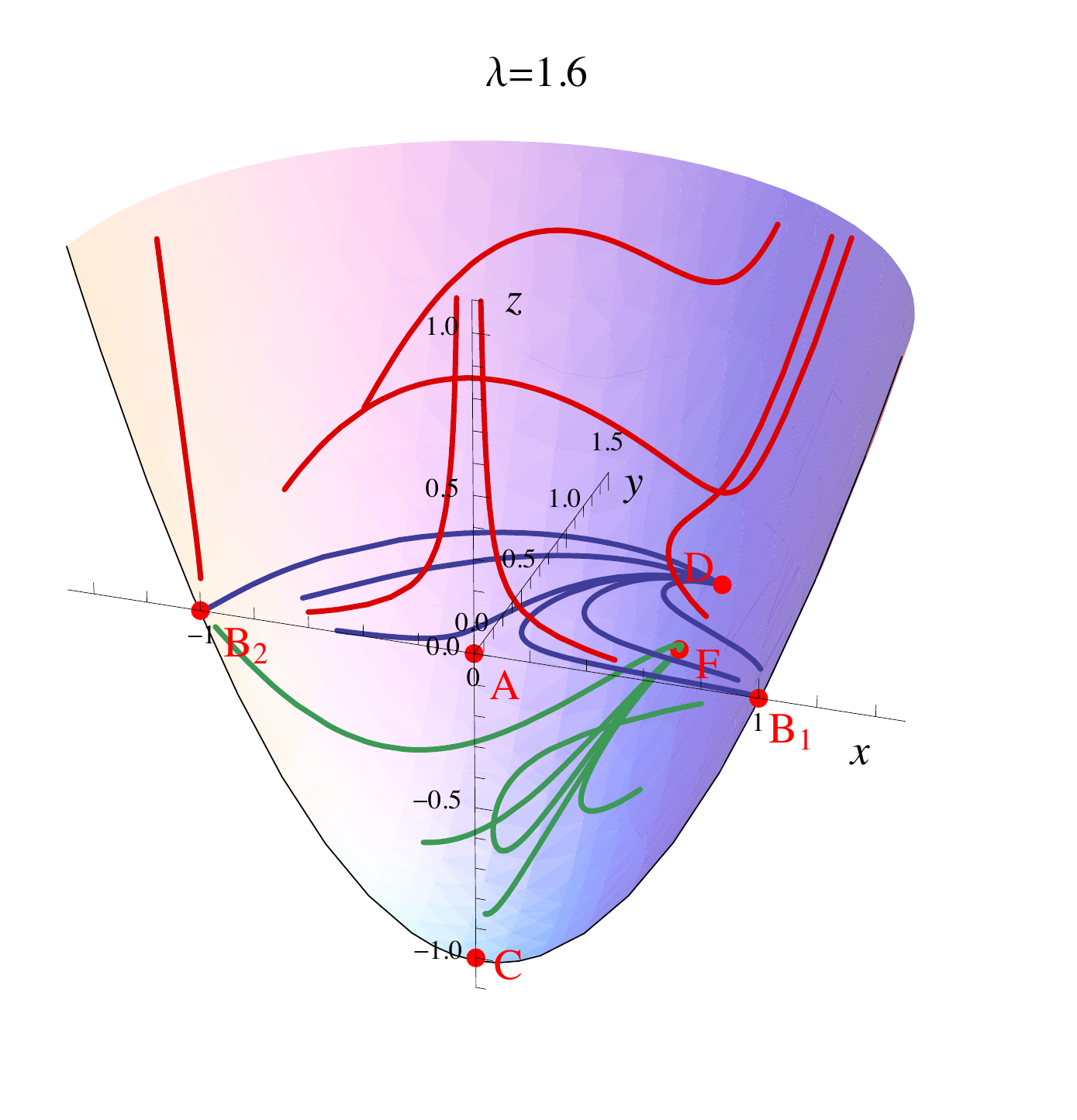}
}
\subfigure{
\includegraphics[scale=0.53]{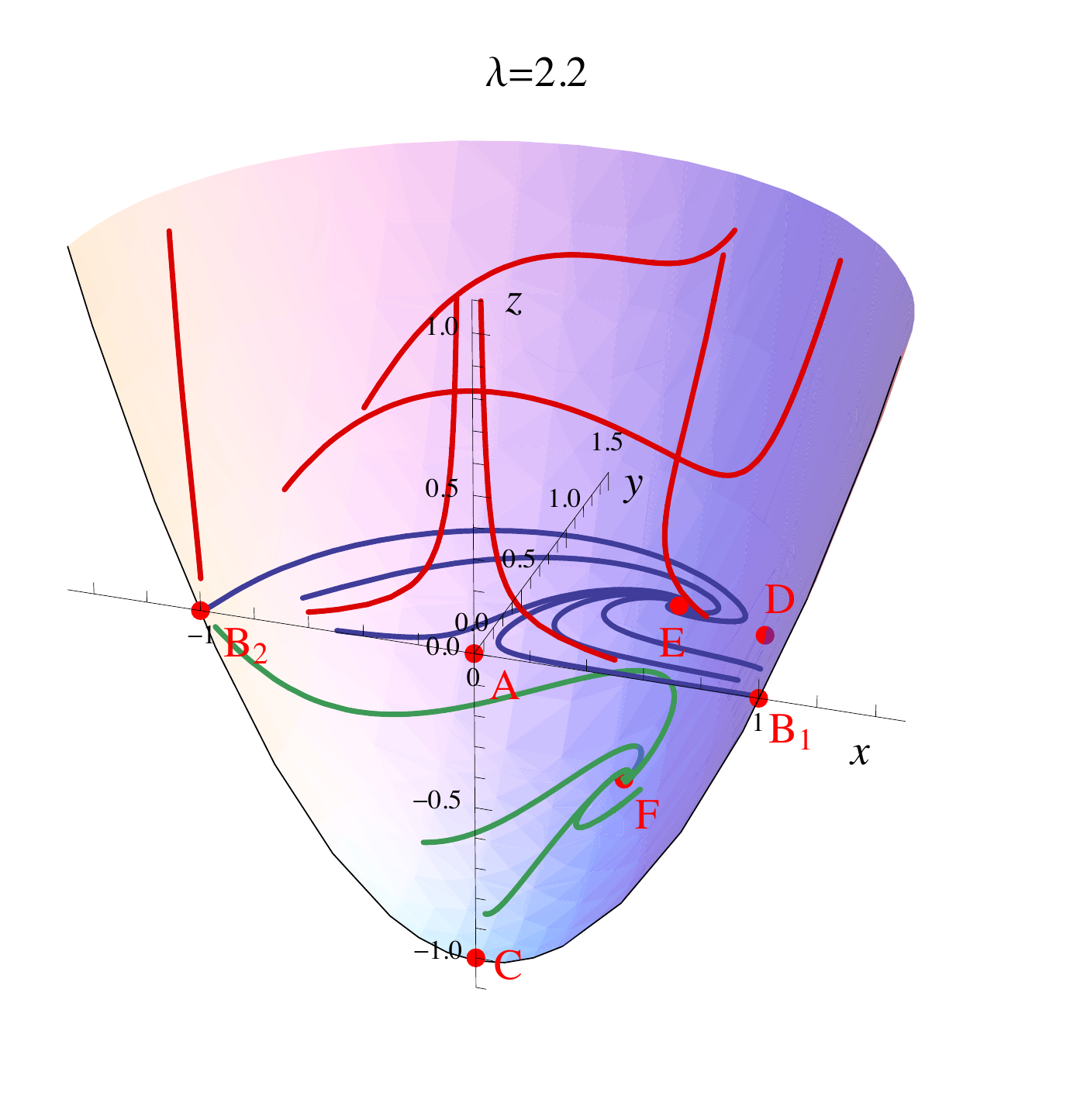}
}
\subfigure{
\includegraphics[scale=0.53]{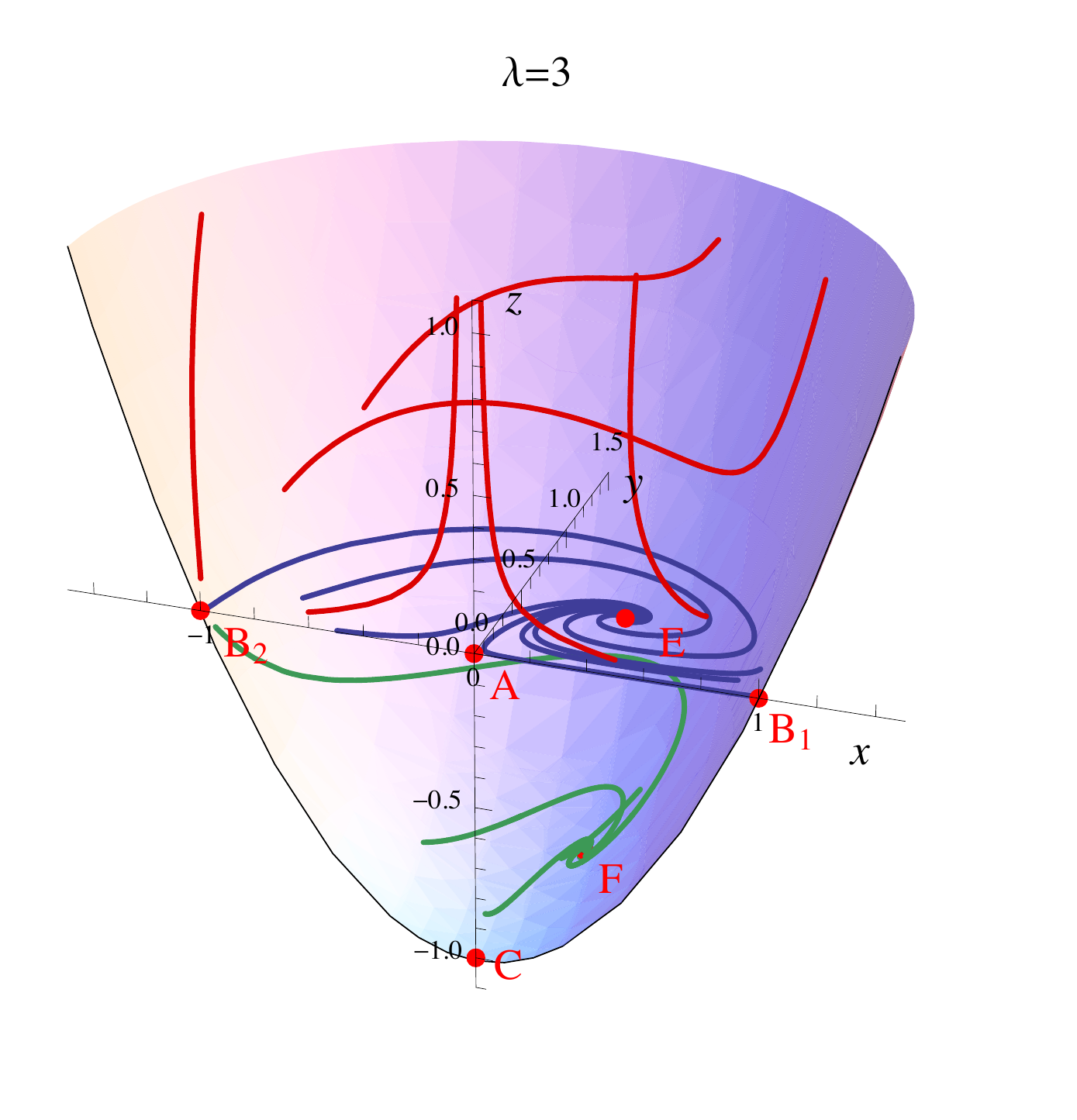}
}
\caption{Phase spaces for four different $\lambda$ and $\wM=0$. Valid phase space is limited by the parabolical surface: all valid trajectories lie above it. Trajectories in the $x-y$ plane (where $z=0$) are shown in blue. Green trajectories lie between this plane and the parabolic surface. Since $z=-\OmK=K/(a^2H^2)$, green trajectories represent part of the phase space with negative curvature. Red trajectories are bounded to the part of the space where $z > 0$ which corresponds to $K>0$. Point \textbf{E} lies under the parabolic surface for the first two cases, so it's not shown in the plots.
\label{pic_3D_phase_space}}
\end{figure}

In this case under consideration the constraint $x^2 + y^2 \leq 1$
holds only in the plane $z=0$. The general constraint is $x^2+y^2-z
\leq 1$ and it defines a parabolic surface in the space of
parameters $x$, $y$, $z$. All trajectories must lie above this
parabolic surface; they are confined either to the plane $z=0$,
where the curvature is zero, or to the part of the space where $z>0$
and $K$ has a positive sign, or to the finite part of the space
between the parabolic surface and the plane $z=0$ with a negative
$z$-component and thus a negative curvature parameter K. Because
this parameter can not change sign in our model, no trajectories can
cross from $z>0$ to $z<0$ or vice-versa.  Figures
(\ref{pic_3D_phase_space}) show solutions for four different case of
parameter $\lambda$. Again, we ignore $\wM$ by setting it to zero.

Inspection of the fixed points in this new system shows that all
fixed points we had before are preserved, when the newly-added
additional $z$ is fixed at $z=0$. In addition, however, there are
now two new fixed points: \textbf{C} and \textbf{F}. The eigenvalues
corresponding to these points are obtained by extending the matrix
(\ref{eq:app:matrix}) to describe three variables; the additional
dimension of the parameter space requires that there will be an
additional eigenvalue. If that eigenvalue is positive, then the
fixed point is repulsive in the new direction and if it is negative,
the fixed point is attractive. For a point to be an attractor, all
the eigenvalues must be negative at that point. If at least one is
positive, then trajectories will be drawn away in at least one
direction.

\begin{description}

\item[Point A] is the matter dominated solution where all energy density is in $\OmM$.
At this point the only contribution to the effective equation of
state parameter comes from the matter sector, so $\weff=\wM$.  The eigenvalues of this point are $[3(\wM-1)/2,  3(\wM +1)/2, (1 + 3 \wM)]$ so the point is a saddle for all values of $\lambda$, as long as $\wM<1$. It is attractive for the trajectories along the $x$-axis
and repulsive for others. In the limit case of initial potential
being exactly $0$ (the trajectories that start exactly on the
$x$-axis) this point is an attractor.

\item[Points $\rm{\textbf{B}}_{\rm{\textbf{1}}}$ and $\rm{\textbf{B}}_{\rm{\textbf{2}}}$] represent solutions in which the universe is dominated by the kinetic energy of the
scalar field: $\Omp=1$. Here the effective equation-of-state
parameter is constant and the scale factor behaves as $a \propto
t^{2/3}$. Eigenvalues for stability are
$[3(1-\wM),\sqrt{3/2}
\lambda \mp 3, 4]$ 
with the minus sign for
$\rm{\textbf{B}}_{\rm{\textbf{1}}}$ and plus for
$\rm{\textbf{B}}_{\rm{\textbf{2}}}$. 
This means that $\rm{\textbf{B}}_{\rm{\textbf{2}}}$ is always repulsive and $\rm{\textbf{B}}_{\rm{\textbf{1}}}$ repulsive for $\lambda <
\sqrt{6}$ and a saddle for $\lambda > \sqrt{6}$. 

\item[Point C]
is new compared to the two-dimensional case and corresponds to a curvature-dominated solution; all the energy
density resides in $\Omega_K$. This solution is trivial, because it
does not involve any matter or the scalar field. The point is a
saddle for all values of parameters; its eigenvalues are $[-2, 1,
-(1 + 3\wM)]$.

\item[Point D] is a solution where the universe is dominated by the scalar field. All energy density is in $\Omp = x^2 + y^2 =1$, with $\Omega_M =0$.
The effective parameter of state is $\weff = \lambda ^2 /3-1$, so the universe is accelerating for $\lambda < \sqrt{2}$. In the limit case $\lambda \rightarrow 0$, this solution corresponds to the de Sitter expansion dominated by the cosmological constant.
This fixed point lies in the $z=0$ plane, at the edge of the half disc and it moves from $(x=0,y=1)$ for $\lambda=0$ to
$(x=1,y=0)$ for $\lambda=\sqrt{6}$. After this value of $\lambda$ it
is not defined anymore. 
The eigenvalues are $[(\lambda ^2-6)/2,\lambda^2-3(\wM-1),\lambda ^2-2]$ which means that the point is attractive for all trajectories with $z \leq 0$ and some with $z > 0$ for $\lambda < \sqrt{2}$. It is also an attractor for trajectories in the plane $z=0$ for $ \sqrt{2} <\lambda < \sqrt{3(\wM+1)}$, and a saddle for $\sqrt{3(\wM+1)}< \lambda < \sqrt{6}$.
The case of $\lambda < \sqrt{2}$ is especially interesting because in this setup even some of the trajectories that start with a positive amount of curvature end up in the inflationary solution (most of the trajectories which start with the positive curvature evolve to a state with infinite $K$, i.e. the universe collapses). This demonstrates that ``closed" does not imply ``finite". 

\item[Point E] is the so-called ``tracking" solution. Both $\Omp$ and $\OmM$ are between $0$ and $1$, and the effective equation-of-state parameter matches that of the matter: $\weff = \wM$. This means that the universe expands as if it were matter dominated and there is no accelerated expansion. This point is, however, interesting because the scalar field is still present; some of the energy density in stored in $\Omp$.\
This fixed point lies in the $z=0$ plane where 
both its coordinates are infinite for $\lambda =0$. Since that is
outside of the half disc, it is not relevant for this analysis. The
point only becomes relevant as  $\lambda$ becomes larger than
$\sqrt{3(\wM+1)}$. At this value \textbf{E} crosses \textbf{D}
and enters the half disc. Analysis of the stability shows that for
$\lambda >\sqrt{3(\wM+1)}$ two of the
eigenvalues are complex conjugates of each other. This means that for the trajectories in the $z=0$ plane this point is a stable spiral. The additional eigenvalue is always positive so all other trajectories are repelled. 

\item[Point F]
The effective equation-of-state parameter in point \textbf{F} is always 
$\weff=-1/3$. This corresponds to the scale factor being linearly proportional to $t$. The matter-energy density vanishes here so the point always lies exactly on the parabolic surface $x^2+y^2-z = 1$. The corresponding eigenvalues are $[-3 \wM-1,-\sqrt{8 \lambda ^4-3 \lambda ^6}/\lambda ^3-1,\sqrt{8 \lambda ^4-3 \lambda ^6}/\lambda ^3-1]$ which results in this point being either a saddle, attractor or a stable spiral, depending on the value of $\lambda$. This is the analogue of the tracking solution (point \textbf{E}) for the case where the universe behaves as if curvature were the only component. This implies $a \propto t$. 
\end{description}

\begin{table}[ht]
\centering
\caption{List of fixed points and theirr properties for the three dimensional case.}
\label{tab:3D}
\begin{tabular}{ |c || c  c  c  || c  c c c|}
\hline
 & $x_*$ & $y_*$ & $z_*$ & $\weff$ & $\Omp$& $\OmM$& $\OmK$\\
\hline
\hline
A & 0 & 0 & 0 & $\wM$& 0 &1 & 0\\
\hline
$\mathrm{B}_1$  & 1 & 0 & 0 & 1 &1&0 &0\\
\hline
$\mathrm{B}_2$  & -1 & 0 & 0 & 1 &1&0&0\\
\hline
C  & 0 & 0 & -1 & undefined &0 &0&1 \\
\hline
D  & $\lambda / \sqrt{6}$ & $\sqrt{1-\lambda ^2 / 6}$ & 0 & $\lambda ^2 /3-1$&1 &0&0\\
\hline
E & $\frac{\sqrt{3/2} ( \wM +1) }{ \lambda} $ & $\frac{\sqrt{3/2( 1-\wM^2)}}{\lambda} $& 0 & $\wM$& $\frac{3(\wM+1)}{\lambda ^2}$ &$1 - \frac{3(\wM+1)}{\lambda ^2}$&0\\
\hline
F  & $\sqrt{2/3}/\lambda $ & $2/(\sqrt{3} \lambda )$ & $2/\lambda ^2-1$ & $-1/3$&$2/\lambda ^2$ &0&$1-2/\lambda ^2$\\
\hline
\end{tabular}
\end{table}

\newpage
\newpage

\noindent
Four qualitatively different cases of $\lambda$ are distinguishable:
\begin{itemize}
\item ($\lambda < \sqrt{2}$)
There are 6 fixed points (all except \textbf{E}, which is outside of
allowed region). Point \textbf{D} is the attractor for all of
$z\leq0$ part of the space and some trajectories in $z>0$.
\textbf{F} has positive $z$-component and is a saddle.

\item ($ \sqrt{2} <\lambda < \sqrt{3(\wM+1)}$)
There are still the same 6 fixed points but the $z$-component of
\textbf{F} is now negative. This point is a stable attractor for all
trajectories with $z<0$. For trajectories that lie in the plane
$z=0$ the attractor is point \textbf{D}, and trajectories with $z>0$
don't converge.

\item ($\sqrt{3(\wM+1)}< \lambda < \sqrt{6}$)
Point \textbf{D} becomes a saddle, \textbf{E} enters the allowed
region and becomes attractive for trajectories in the plane, while
trajectories with $z>0$ diverge. Two of the eigenvalues of point
\textbf{F} are now complex conjugates of each other while the third
is negative, so this point is a stable spiral in three dimensions.

\item ($\sqrt{6}< \lambda$) At $\lambda= \sqrt{6}$
The  significance of this last case is that point \textbf{D} merges
with $\rm{\textbf{B}}_{\rm{\textbf{1}}}$ and disappears.
$\rm{\textbf{B}}_{\rm{\textbf{1}}}$ is now a saddle point in the
plane.
\end{itemize}

\subsection{Evolution of the Equation of State}
\begin{figure}[ht!]
\centering
\subfigure{
\includegraphics[scale=0.53]{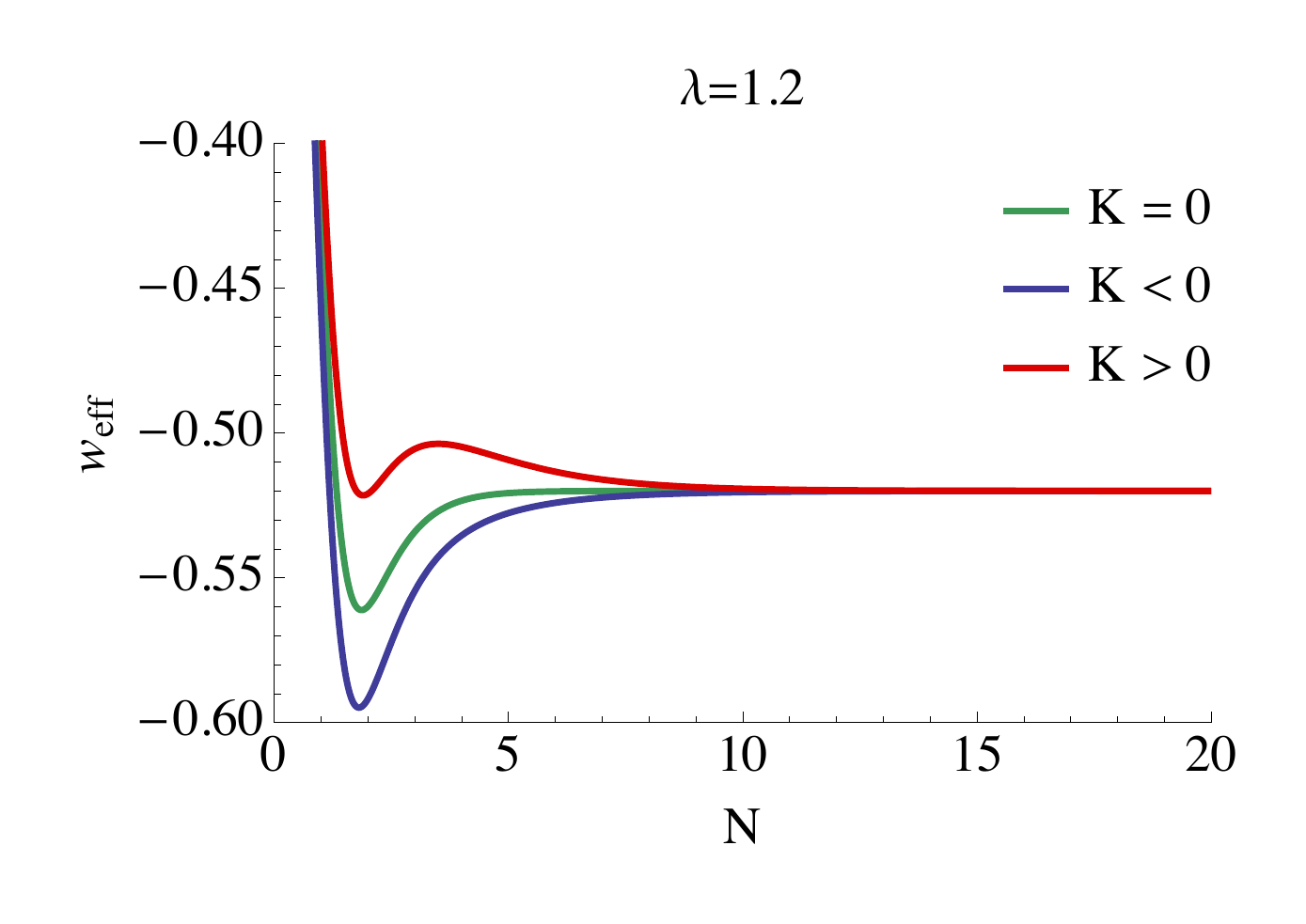}
}
\subfigure{
\includegraphics[scale=0.53]{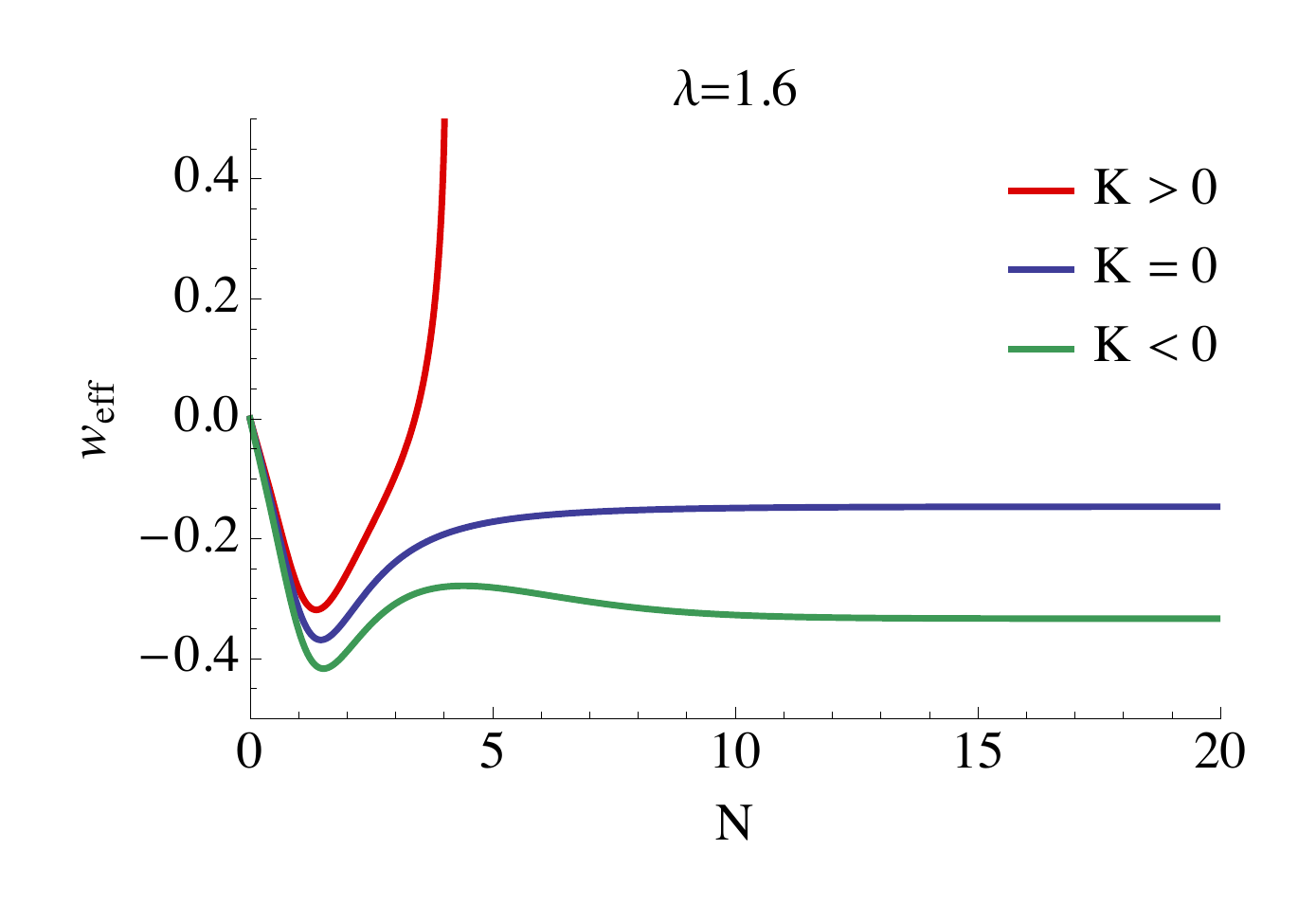}
}
\subfigure{
\includegraphics[scale=0.53]{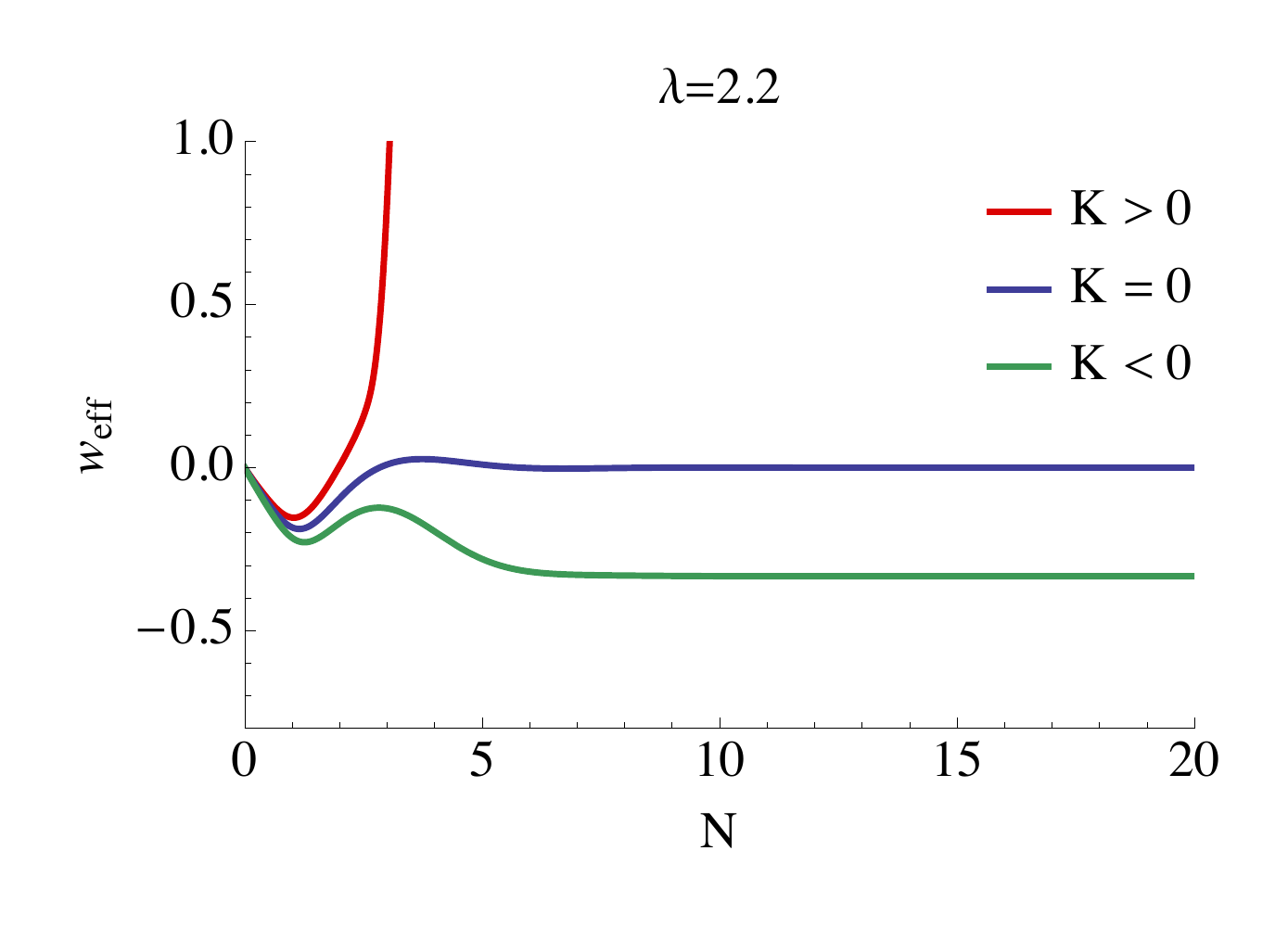}
}
\subfigure{
\includegraphics[scale=0.53]{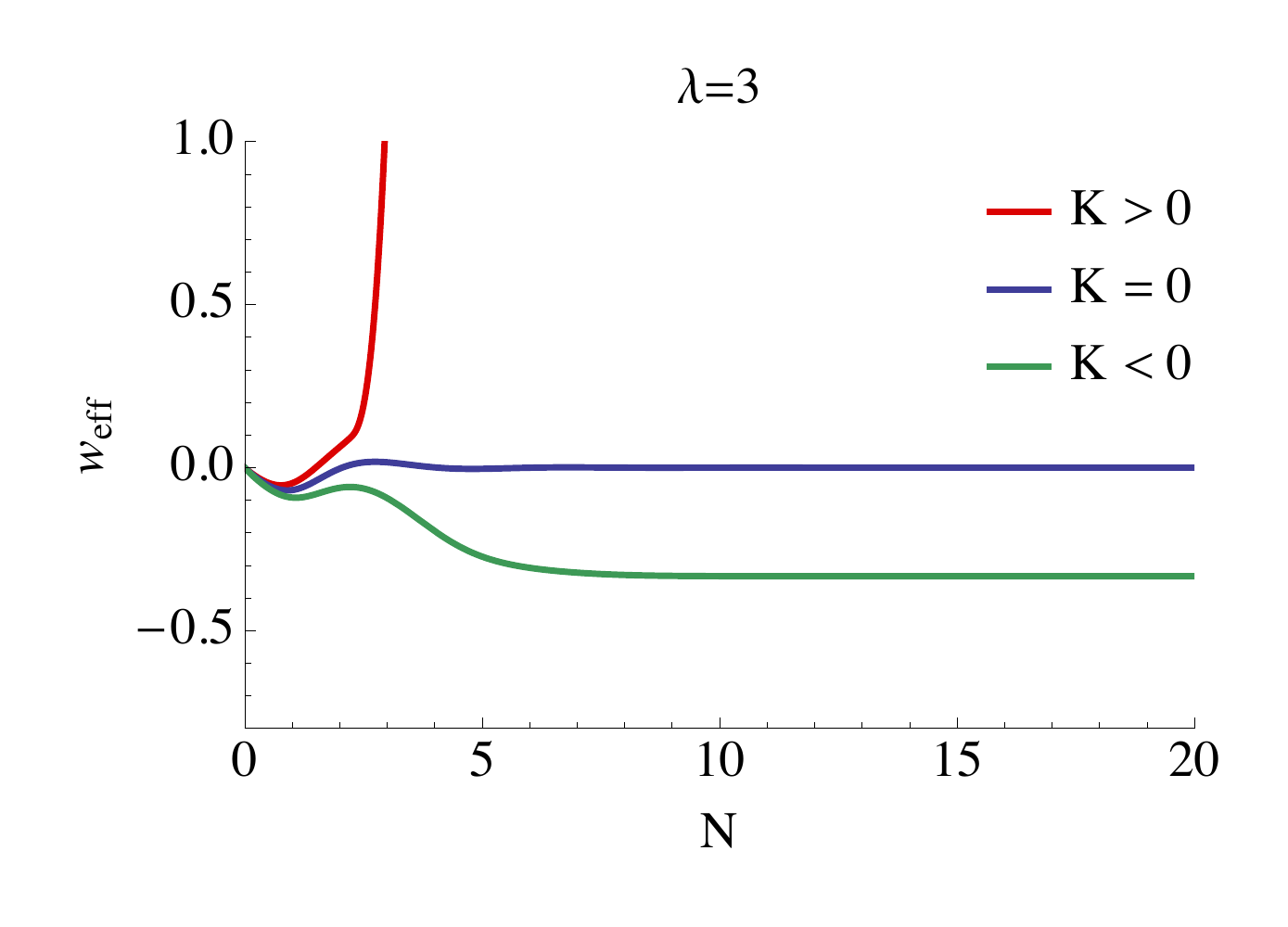}
}
\caption{Evolution of equation-of-state parameter over $20$ e-foldings (N) for four different cases of $\lambda$.\label{pic_weffEvolution}}
\end{figure}

\noindent
For the same cases of $\lambda$ we plot trajectories that correspond
to initial conditions that give flat, open and closed universe
models. These are shown in figures (\ref{pic_weffEvolution}). Note
that we used the same initial conditions for all four cases. In the
first case all curves converge. For all other values of $\lambda$
the equation-of-state parameter diverges for $K>0$. For $K<0$
trajectories in phase space are drawn towards point \textbf{F},
where $\weff=-1/3$. For $K=0$ they can either end up in the point
\textbf{D} (first two values of $\lambda$) or \textbf{E} (third and
fourth case of $\lambda$).

\newpage
\section{Conclusions}
\noindent
We explored a single-field quintessence model with exponential
potential and canonical kinetic term in presence of a FLRW metric
and overall (positive or negative) spatial curvature. 
In the Friedmann Equations, the term with spatial curvature is
dynamical, so we introduced a new variable and generalized
two-dimensional case that holds for a flat geometry to three
dimensions.

In comparison to the model without curvature there are two new fixed
points. These have been noticed before (e.g. \cite{burdbar88}) but
their behaviour is clearer when viewed in terms of the
parametrization we use. One of them is trivial, in that it corresponds to a universe
where entire energy density is dominated by the curvature. This
fixed point is not attractive. The other fixed point corresponds to
a universe where energy density is a combination of curvature and
scalar field (but no matter). The ratio between them depends on the
parameter of the exponential potential. This fixed point is
interesting, because it is attractive for all trajectories with
negative curvature term for $\lambda > \sqrt{2}$. We have thus
established that there is a natural configuration of this system
that corresponds to a specific value of $w$. Although we know that the
energy density of the Universe today is almost critical (so there is
very little or no curvature), this fixed point might still be
interesting for the cases where $\lambda$ is just above $\sqrt{2}$
and so energy density is dominated by the scalar field. 

The only problem with relating this solution to our observed reality is that
equation-of-state parameter in this point is $-1/3$, but we know
that the present value is very close to $-1$. Nonetheless, the attractor solution \textbf{D} is an accelerating solution for flat, negatively curved and some positively curved cases. The corresponding equation-of-state parameter is close to the value $w=-1$ in the limit $\lambda \rightarrow 0$, but for any value of $\lambda < \sqrt{2}$ it corresponds to $w<-1/3$ so this is at least qualitatively applicable to dark energy. 

We have only investigated one particular and rather simple model, but this sort of curvature inclusive analysis is applicable to other,
potentially more complicated models that might exhibit a much richer
dynamical interplay. Cases which might prove amenable to further
study on these lines would be scalar fields with a more general potential \cite{0904.0877}, non-canonical
terms, with multiple scalar fields, with globally anisotropic metrics
(i.e. the Bianchi models) \cite{gr-qc/1308.1658}, e.g. the Kantowski-Sachs model \cite{gr-qc/0004060}, and models based on exact inhomogeneous
cosmologies. Such models would of course introduce more than one
extra parameter, so the resulting phase portraits would involve more
than three dimensions and their analysis would entail considerably
greater complexity.

\section*{Acknowledgements}
 \noindent
PC gratefully acknowledges support from the UK Science and Technology Facilities Council (STFC) under grant reference ST/L000652/1. MG would like to thank Nicola Tamanini for useful comments. 

\appendix\label{app:two-variables}
\subsection{Two dynamical variables and stability of fixed points}
\noindent
After introducing new variables as in Equation (\ref{eq:variables2D}) for a general potential $V(\phi)$ and using the Klein-Gordon Equation (\ref{eq:continuity:scalar}), the derivatives with respect to $N$ are: 
\begin{subequations} 
\begin{equation}\label{eq:app:dinsysraw}
\frac{dx}{dN} = - \frac{\dot{\phi}}{\sqrt{6}\MPl H} \left( 3+ \frac{\dot{H}}{H^2}\right)- \frac{V_{,\phi}}{\sqrt{6}\MPl H^2} 
\end{equation}
\begin{equation}
\frac{dy}{dN} =  \frac{V_{,\phi} \; \dot{\phi}}{2\sqrt{3}\MPl \sqrt{V} H^2} - \frac{\sqrt{V}}{\sqrt{3}\MPl H}\frac{\dot{H}}{H^2}.
\end{equation}
\end{subequations}
Defining parameter $\lambda$ as $\lambda=-\MPl V_{,\phi}/V$ is especially convenient because this parameter is constant for the exponential potential. 
Using this potential and eliminating $\OmM$ with the use of
constraint (\ref{eq:constraint}), the ratio of the Friedmann
Equations $\dot{H}/H^2$ can be expressed, in terms of $x$ and $y$,
as:
\begin{equation}
\frac{\dot{H}}{H^2}=\frac{3}{2} \left(\wM \left(x^2+y^2-1\right)-x^2+y^2-1\right),
\label{eq:app:ratio}
\end{equation}
where we define $\wM$ to be the combined equation-of-state parameter
for a fluid component taken to describe both matter and radiation:
\begin{equation}
\wM=\frac{\rhom \wm + \rhor \wr}{\rhom +\rhor}.
\end{equation}
Derivatives of $x$ and $y$ are hence expressed, for the exponential potential as (\ref{eq:derivatives}). This provides a useful way to express the
evolution of both $\wphi$ and $\Omp$:
\begin{subequations}
\begin{equation}
\wphi ' = (\wphi-1) \left(- \lambda  \sqrt{3(1+\wphi) \Omp}+3(1+ \wphi)\right)
\end{equation}
\begin{equation}
\Omp ' = 3 (\Omp-1) \Omp (\wphi-\wM),
\end{equation}
\end{subequations}
where the prime stands for differentiation with respect to $N$.
Additionally, by defining $\Gamma=V V_{,\phi\phi} / V_{,\phi}^2$ we
can obtain the derivative of $\lambda$:
\begin{equation}
\lambda '= -\lambda^2(\Gamma -1) \sqrt{3 (1+\wphi)\Omp}.
\end{equation}
For the exponential potential $\lambda$
is constant, so $\lambda'$ does not carry any additional information in this case.
In the fixed points the effective equation-of-state parameter,
defined as
\begin{equation}
\weff=\frac{P_{\phi}+\PM}{\rho_{\phi}+\rhoM}=x^2-y^2+\wM(1-x^2-y^2)
\end{equation}
is constant so we can easily integrate Equation (\ref{eq:app:ratio})
to see how scale factor changes with time:\begin{equation}
a \propto t^{\frac{2}{3(1+\weff)}}.
\end{equation}
Note that this only holds if the system is at one of the fixed
points. In other cases Equation (\ref{eq:app:ratio}) has to be
integrated numerically because its right-hand side is not constant.
However, it is still sensible to define an equation-of-state
parameter $\weff$ even outside these points. For $\weff=-1/3$ the
scale factor will have a linear dependance on time; values larger or
smaller than that will give decelerated and accelerated expansion,
respectively.

We investigate the stability of the fixed points by finding
eigenvalues of the matrix, evaluated at the fixed point $(x_*,y_*)$:
\begin{equation}
M=\left[\begin{array}{cc} \frac{ \partial f_x}{\partial x} &
\frac{\partial f_x}{\partial y}   \\  \frac{\partial f_y}{\partial
x} & \frac{\partial f_y}{\partial y} \end{array} \right]_{x_*,y_*},
\label{eq:app:matrix}
\end{equation}
where $f_i=di/dN$ stands for the terms on the right-hand side of
Equations (\ref{eq:derivatives}). A fixed point is stable (attractive)
if both eigenvalues are negative, non-stable (repulsive) if they are
positive and a saddle (attractive in one direction and repulsive in
others) if they have the opposite sign. If the two eigenvalues are
complex conjugates of each other, then the fixed point is a stable
spiral.

In the case of two variables there are five such critical points; their coordinates and values of
the associated physical parameters are listed in Table
(\ref{tab:2D}). Their existence and properties depend on the values
of $\lambda$ and $\wM$.

\clearpage
\bibliographystyle{utcaps}
\bibliography{references}

\end{document}